\documentclass[twocolumn,aps,superscriptaddress]{revtex4-2}

\usepackage[utf8]{inputenc}
\usepackage{amsmath,amssymb,graphicx}
\usepackage{braket}
\usepackage{hyperref}
\usepackage{xcolor}
\usepackage{dcolumn}
\usepackage{mathtools}
\usepackage{tabularx}

\usepackage{booktabs}   
\usepackage{pifont}     
\usepackage{makecell}   
\newcommand{\cmark}{\ding{51}}
\newcommand{\xmark}{\ding{55}}

\usepackage{tikz}
\usetikzlibrary{arrows.meta,positioning,calc,fit,shapes.geometric,shapes.misc}

\newcommand{\pmark}{(\kern-0.2em$\sim$\kern-0.2em)}
\begin{document}

\title{Quantum Erasure Imaging: Complementary Modalities from Delayed-Choice Erasure}

\author{Sean D. Huver}
\affiliation{NVIDIA}

\author{Sanjaya Lohani}
\affiliation{Southern Methodist University}


\begin{abstract}
Quantum Erasure Imaging (QEI) turns delayed-choice erasure into a practical imaging protocol. Entangled photon pairs encode two classical modalities, absorption \(T(x,y)\) and a phase-sensitive cosine quadrature of \(\phi(x,y)\), reconstructed from a single run of time-tagged coincidences by retrospective sorting on a remote ancilla. Measuring the ancilla in H/V yields \(T\) via which-path information; D/A yields interference visibility $\propto \frac{2\sqrt{T}}{T+1}\cos\phi$; and a rotated orthonormal analyzer continuously trades between them. We derive balanced two-port estimators whose denominators are analyzer independent (completeness / no signaling), together with Fisher information (FI) and Cram\'er--Rao bounds (CRBs) that establish an equivalence to time division under labeled randomization. The advantages of QEI are operational: single-run acquisition, perfect co-registration, and remote / delayed mode choice. We illustrate the protocol with Monte-Carlo simulations and open source our code.
\end{abstract}

\maketitle

\section{Introduction}

QEI turns delayed-choice quantum erasure into an imaging protocol \cite{Wheeler1978,Scully1982,Kim2000,walborn2002double}: from a single labeled acquisition of time-tagged coincidences, a remote measurement on an entangled ancilla retrospectively sorts the event stream into complementary imaging channels. The which-path channel reconstructs absorption \(T(x,y)\)~\cite{bagan2016relations}, while the erasure channel reconstructs a co-registered phase-sensitive visibility proportional to \(\cos\phi(x,y)\). Rotating the ancilla analyzer continuously tunes the relative weight of which-path information and interference contrast.

The protocol does not circumvent complementarity: the analyzer angle sets predictability and visibility in accordance with the standard bound \cite{Englert1996, jacques2008delayed} (saturating it near \(T\simeq 1\)). The complete protocol delivers single-run acquisition of co-registered complementary observables with the object arm unchanged and basis choices that can be randomized or delayed within the same time-tagged dataset.

Quantum‑eraser experiments, including delayed‑choice variants, have long established the role of context‑dependent interference~\cite{Scully1982,Kim2000, lin2026fringe,li2021delayed, imran2018doubly, kim2023observations,yu2026experimental,ionicioiu2011proposal,kaiser2012entanglement} , and QEI draws directly on this foundation by repurposing specifically the underlying mechanism for {image formation} with explicit $T/\phi$ estimators from one dataset. Additionally, Ghost imaging retrieves spatial structure through correlations~\cite{Pittman1995,Erkmen2010,Bennink2002, tan2022single, schiano2024engineering,liu2018fast}, however, QEI departs from this approach by placing the object inside an interferometer of the imaging arm and using the ancilla basis to decode the imaging modalities. Furthermore, imaging with undetected photons, in contrast, relies on induced coherence~\cite{Lemos2014}, whereas QEI uses conditional post‑selection on an entangled partner to switch seamlessly between absorption and phase reconstructions within the same experimental run. Prior demonstrations of two‑photon imaging quantum erasers and random delayed‑choice implementations have likewise shown wave–particle behavior in imaging geometries~\cite{scarcelli2007, wang2019quantum}, and QEI extends this trajectory by integrating these ideas into a unified, operationally efficient imaging protocol.


\begin{figure*}[t]
\centering
\includegraphics[width=.95\linewidth]{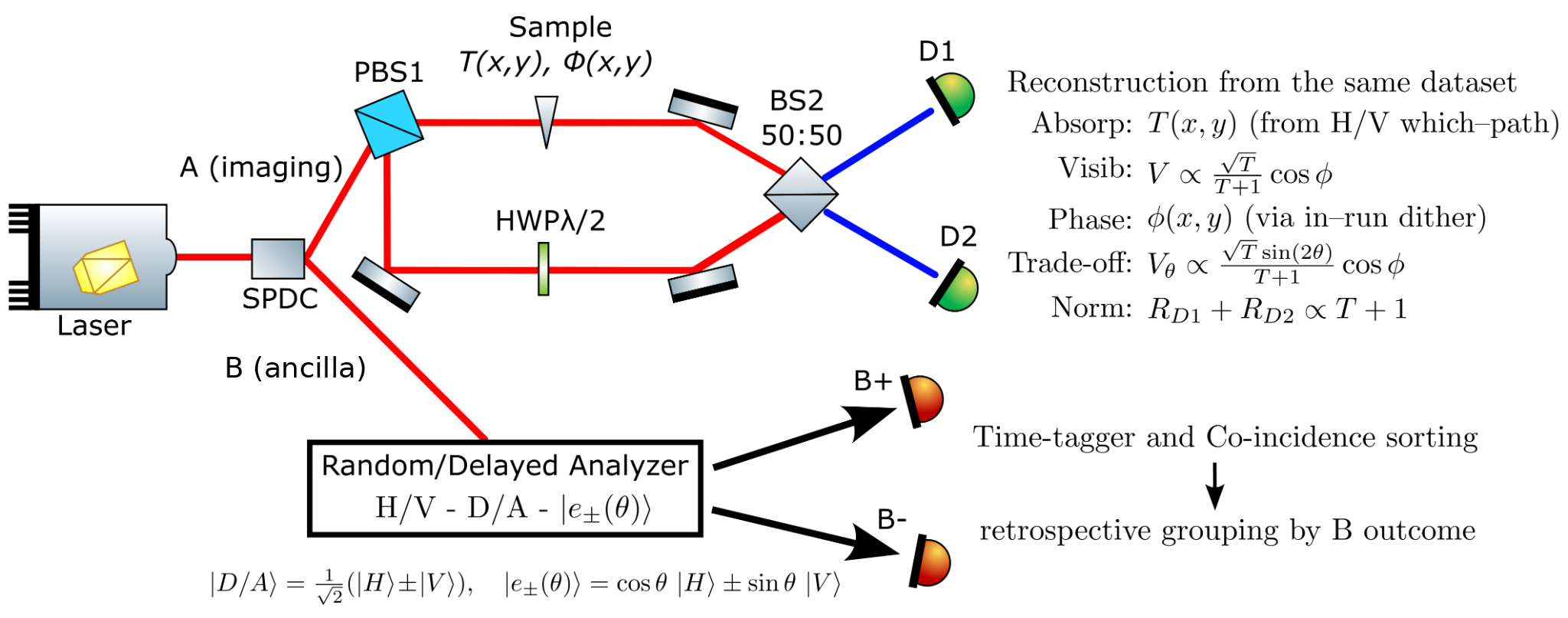}
\caption{\textbf{QEI schematic.} The imaging photon $A$ traverses a modified Mach--Zehnder:
PBS1 encodes which--path on polarization; the object in Arm~1 imparts $T(x,y),\phi(x,y)$; a HWP equalizes polarization; and BS2 recombines to two ports (D1,D2).
The ancilla photon $B$ is measured in a random/delayed analyzer (H/V, D/A, or a rotated orthonormal basis $\{\ket{e_{\pm}(\theta)}\}$).
All events are time--tagged and retrospectively sorted by $B$'s outcome to reconstruct co--registered absorption and phase images from a single run.
Two--port normalization makes the denominator analyzer--independent (completeness/no--signaling).}
\label{fig:qei-schematic}
\end{figure*}

Moreover, we expand the prior work in three ways. First, we obtain co‑registered dual‑modality ($T,\phi$) reconstructions from the same time‑tagged acquisition, partitioned into labeled basis subsets. Second, we introduce a continuous, orthonormal analyzer sweep for 2D images, together with a two‑port normalized estimator whose denominator is analyzer‑independent, ensuring completeness and no‑signaling. Third, we provide closed‑form estimators and FI/CRBs \cite{Helstrom1976}, along with a precise demonstration of Fisher‑Information‑Matrix equivalence to time‑division, clarifying that QEI’s advantages are operational rather than per‑photon informational. A comparison of QEI with related quantum and entanglement‑based imaging methods is provided in Table \ref{tab:method_comparison} in the Appendix.

\section{Theoretical Framework}
We prepare polarization-entangled pairs via type-II spontaneous parametric down-conversion (SPDC) \cite{Kwiat1995}:
\begin{equation}
\ket{\Psi_{0}}=\tfrac{1}{\sqrt{2}}\!\left(\ket{H}_A\ket{H}_B+\ket{V}_A\ket{V}_B\right).
\end{equation}
PBS1 maps $\ket{H}_A\!\to$ Arm~1 (sample), $\ket{V}_A\!\to$ Arm~2 (reference), creating path--polarization entanglement:
\begin{equation}
\ket{\Psi_{1}}=\tfrac{1}{\sqrt{2}}\!\left(\ket{\mathrm{H}}_{Arm1_A}\ket{H}_B+\ket{\mathrm{V}}_{Arm2_A}\ket{V}_B\right).
\end{equation}
The sample imparts transmission $T(x,y)$ and phase $\phi(x,y)$ in Arm~1. A half-wave plate in Arm~2 rotates $\ket{V}\!\to\!\ket{H}$ so both paths share polarization at recombination. Note that for absorbing samples ($T<1$), this unnormalized expression implicitly conditions on the photon not being lost to vacuum:
\begin{equation}
\ket{\Psi_{2}}=\tfrac{1}{\sqrt{2}}\!\left(\sqrt{T}e^{i\phi}\ket{\mathrm{H}}_{Arm1_A}\ket{H}_B+\ket{\mathrm{H}}_{Arm2_A}\ket{V}_B\right).
\end{equation}
BS2 (50/50) maps
\begin{align}
\ket{\mathrm{H}}_{Arm1_A} &\rightarrow \tfrac{1}{\sqrt{2}}\!\left(\ket{D1}_A+\ket{D2}_A\right),\\
\ket{\mathrm{H}}_{Arm2_A} &\rightarrow \tfrac{1}{\sqrt{2}}\!\left(\ket{D1}_A-\ket{D2}_A\right),
\end{align}
yielding
\begin{equation}
\label{eq:final_state}
\begin{split}
    \ket{\Psi_{3}} = \tfrac{1}{2}\! \Big[ & \ket{D1}_A(\sqrt{T}e^{i\phi}\ket{H}_B+\ket{V}_B) \\
    & + \ket{D2}_A(\sqrt{T}e^{i\phi}\ket{H}_B-\ket{V}_B) \Big].
\end{split}
\end{equation}

\subsection{Reconstruction in Canonical Bases}
Measuring B in H/V has total which-path information (WPI) of the photon in A, and we find at either port,
\begin{equation}
R(D1,H)\propto T,\quad R(D1,V)\propto 1\quad (\text{and similarly for D2}),
\end{equation}
so the two-port absorption estimator is
\begin{equation}
\widehat{T}(x,y)=\frac{R(D1,H)+R(D2,H)}{R(D1,V)+R(D2,V)}.
\label{eq:T_estimator_twoports}
\end{equation}

\begin{figure}[b]
\centering
\includegraphics[width=\linewidth]{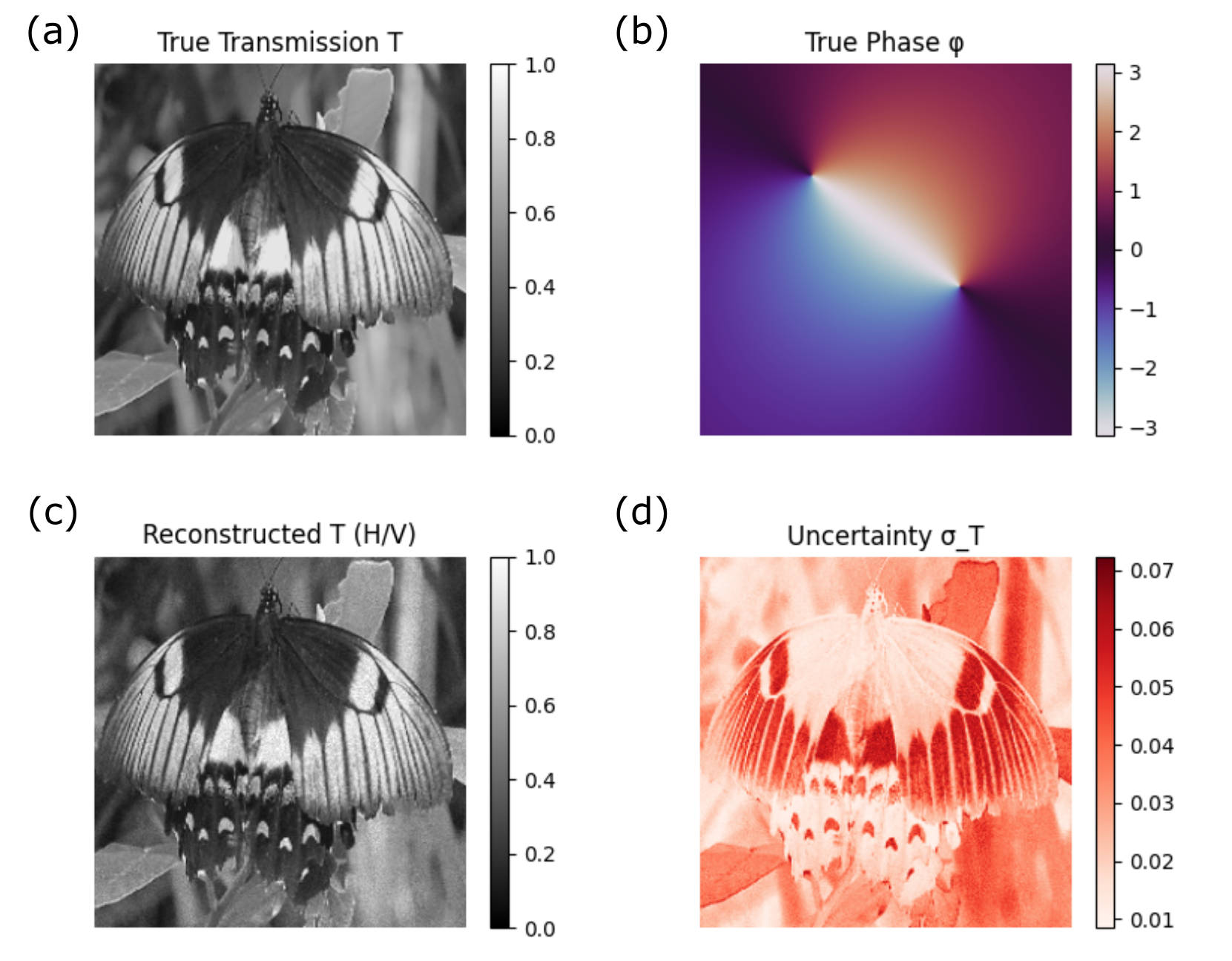}
\caption{Illustration of the reconstruction of $T(x,y)$ using $H/V$ polarization bases. (a) Image representing True $T(x,y)$. (b) A vortex phase map associated with the sample. Measuring $B$: (c) Reconstructed $T(x,y)$. (d) Estimated uncertainty in the reconstructed $T(x,y)$.}
\label{fig:T-vs-recons-T}
\end{figure}

\begin{figure}[t]
\centering
\includegraphics[width=\linewidth]{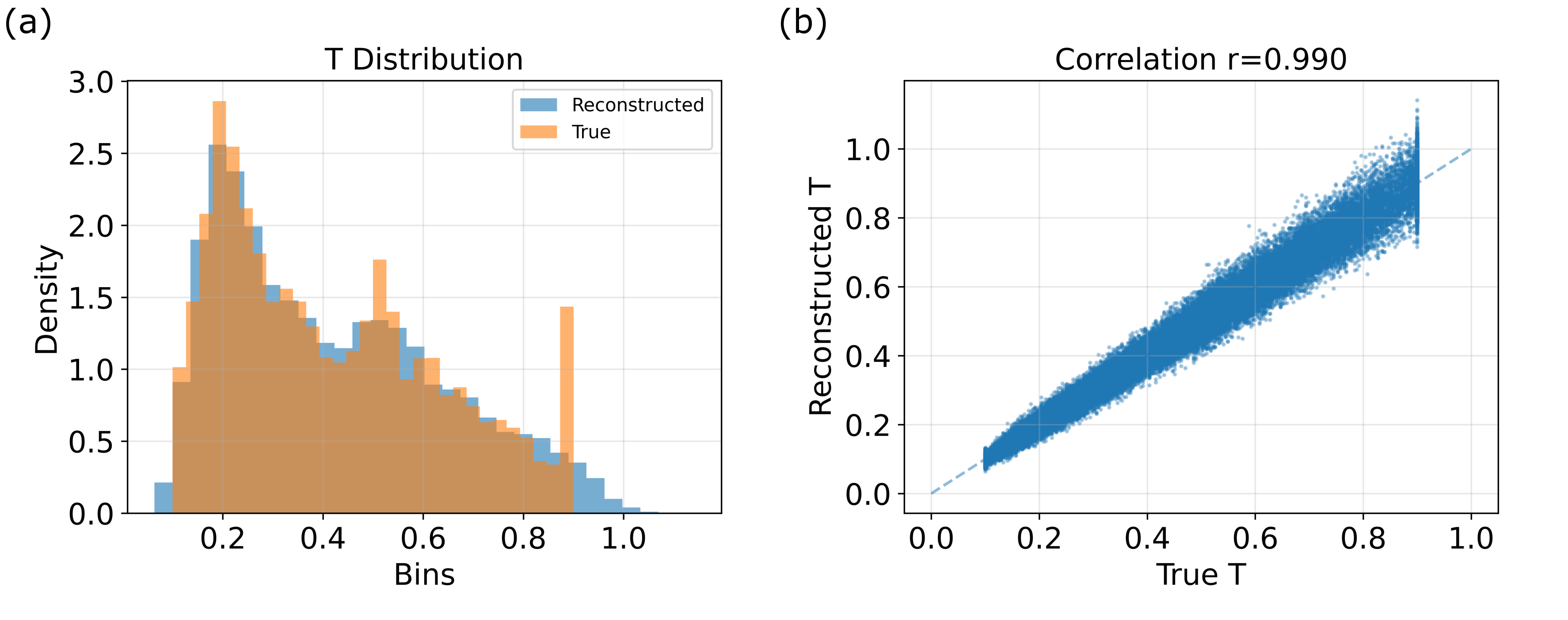}
\caption{Measuring $B$ on $H/V$ bases. (a) $T(x,y)$ Distribution: brown and blue histograms, respectively, represent the true $T(x,y)$ and reconstructed values of $T(x,y)$. (b) Correlation coefficient between the true $T$ and reconstructed $T$ at a simulated $1000$ photons per pixel.}
\label{fig:T-vh-r-coeff}
\end{figure}

To demonstrate the reconstruction process, we begin with an image~\cite{phucthai2022butterfly} representing the true transmission function $T(x,y)$, shown in Fig. \ref{fig:T-vs-recons-T} (a). After that, a vortex phase map, $\phi(x,y)$, is simulated and presented in Fig. \ref{fig:T-vs-recons-T} (b). A simulated single acquisition with 1000 photons per pixel is then performed by measuring B in H/V bases using our open-source Monte-Carlo framework \cite{qei_github}. Using the coincidence counts recorded at detectors D1 and D2, the transmission function $T(x,y)$ is reconstructed. The resulting reconstruction and its associated uncertainty are shown in Fig. \ref{fig:T-vs-recons-T} (c) and Fig. \ref{fig:T-vs-recons-T} (d), respectively. Moreover, we can clearly find a significant overlap between the true and reconstructed $T(x,y)$, as indicated in the histogram plot, Fig. \ref{fig:T-vh-r-coeff} (a). We also find the correlation coefficient $r=0.990$ between the true and reconstructed $T(x,y)$ as shown in Fig. \ref{fig:T-vh-r-coeff} (b).

Measuring B in D/A with $\ket{D/A}=(\ket{H}\pm\ket{V})/\sqrt{2}$ erases WPI, and yields, at D1,
\begin{align}
R(D1,D)&\propto T+1+2\sqrt{T}\cos\phi,\\
R(D1,A)&\propto T+1-2\sqrt{T}\cos\phi,
\end{align}
and the opposite fringe at D2. The two-port normalized differential (equal to single-port visibility) is

\begin{equation}
\label{eq:V_DA}
\begin{split}
    V(x,y) &= \frac{[R(D1,D)-R(D1,A)]-[R(D2,D)-R(D2,A)]}{[R(D1,D)+R(D1,A)]+[R(D2,D)+R(D2,A)]} \\
           &= \frac{2\sqrt{T(x,y)}}{T(x,y)+1}\cos\phi(x,y).
\end{split}
\end{equation}

The visibility of a simulated single acquisition for the example image is illustrated in Fig. \ref{fig:visibility}. The resulting contrast of interference pattern while measuring B in D/A bases and its associated uncertainty are shown in Fig. \ref{fig:visibility} (a) and (b), respectively.

\subsection{Intermediate Analyzer}
We can also introduce an orthonormal rotation to continuously tune which-path information versus interference. We rotate B's polarization in the following orthonormal basis:
\begin{subequations}
\label{eq:orthobasis}
\begin{align}
    \ket{e_1(\theta)} &= \cos\theta\,\ket{H}+\sin\theta\,\ket{V}, \\
    \ket{e_2(\theta)} &= -\sin\theta\,\ket{H}+\cos\theta\,\ket{V}.
\end{align}
\end{subequations}
Projecting Eq.~\eqref{eq:final_state} onto $\{\ket{e_{1,2}}\}_B$, the D1 intensities are
\begin{align}
R(D1,e_1)&\propto T\cos^{2}\theta+\sin^{2}\theta+2\sqrt{T}\sin\theta\cos\theta\cos\phi,\label{eq:D1e1}\\
R(D1,e_2)&\propto T\sin^{2}\theta+\cos^{2}\theta-2\sqrt{T}\sin\theta\cos\theta\cos\phi,\label{eq:D1e2}
\end{align}

with the interference term reversing sign at D2. Summing over $\ket{e_{1,2}}$ on a given port gives $T+1$ (up to normalization), independent of $\theta$, as required by completeness/no-signaling:
\begin{subequations}
\label{eq:singles_nosignal}
\begin{align}
    R(D1,e_1)+R(D1,e_2) &\propto T+1, \\
    R(D2,e_1)+R(D2,e_2) &\propto T+1.
\end{align}
\end{subequations}
Using both ports, the normalized differential (generalizing Eq.~\eqref{eq:V_DA}) is
\begin{equation}
V_{\theta}(x,y)=\frac{2\sqrt{T(x,y)}\,\sin(2\theta)}{T(x,y)+1}\,\cos\phi(x,y),
\label{eq:V_theta}
\end{equation}
which reduces to Eq.~\eqref{eq:V_DA} at $\theta=\pi/4$ and vanishes at $\theta=0,\pi/2$. Notably, for the two-port estimator the {denominator is $\theta$-independent}.

\section{Estimator Statistics and Cram\'er--Rao Bounds}
Let $N$ be the expected {coincidences per pixel} in one run; a fraction $\alpha$ is measured in H/V (WPI) and $\beta=1-\alpha$ in D/A or a rotated basis.

\subsection{Absorption ($T$) from WPI}
Pooling both ports gives two outcomes $\{H,V\}$ with probabilities $p_H=T/(T+1)$, $p_V=1/(T+1)$. With $S_T=\alpha N$ WPI-tagged coincidences,
\begin{align}
\mathrm{Var}\!\left[\widehat{T}\right] &\simeq \frac{T(1+T)^{2}}{S_T}, \qquad
\mathrm{SNR}_T \simeq \frac{\sqrt{S_T\,T}}{1+T}.
\label{eq:Tvar}
\end{align}
This saturates the CRB \cite{Helstrom1976} for WPI-tagged events.

\begin{figure}[b]
\centering
\includegraphics[width=\linewidth]{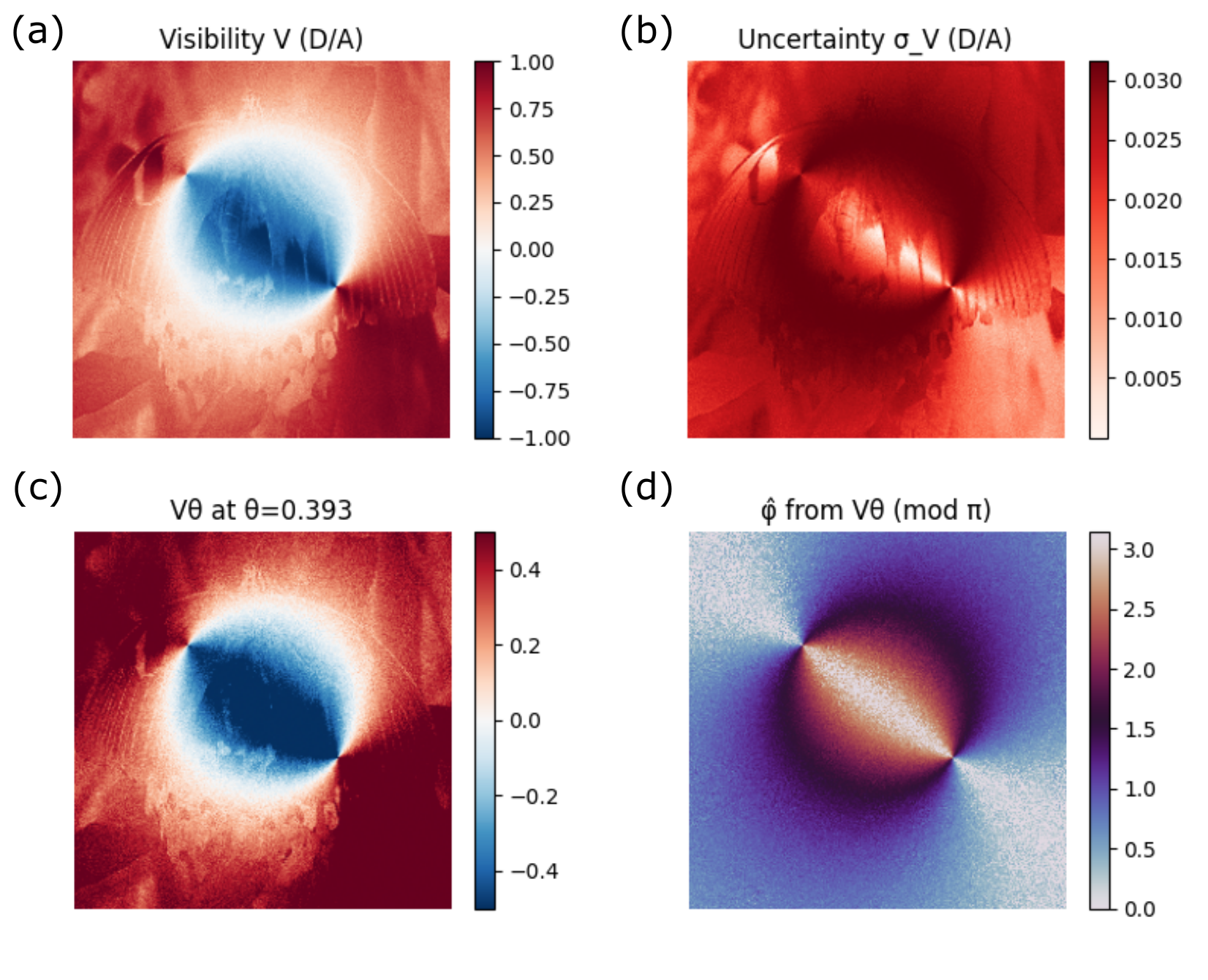}
\caption{Measuring $B$ on $D/A$ bases: (a) Two-port normalized differential corresponding to the given image $T(x,y)$. (b) Estimated uncertainty in the visibility derived from coincidence counting statistics. Measuring $B$ on rotational $|e_{\pm}(\theta)\rangle$ bases: (c) Visibility evaluated at $\theta = \pi/8$. (d) Reconstruction of phase $\phi$ (modulo $\pi$) based on the visibility shown in panel (c).}
\label{fig:visibility}
\end{figure}

\subsection{Phase ($\phi$) from D/A}
Using both ports with $S_\phi=\beta N$ erasure-tagged coincidences, the per-event FI for $\phi$ is
\begin{equation}
\mathcal{I}_{\phi}^{(D/A)}
= \frac{4T\sin^{2}\phi}{(T+1)^{2}-4T\cos^{2}\phi}\!,
\label{eq:IF_phi_DA}
\end{equation}
and the CRB is $\mathrm{Var}[\widehat{\phi}]\ge 1/(S_\phi\,\mathcal{I}_{\phi}^{(D/A)})$. At $T=1$, $\mathcal{I}_{\phi}^{(D/A)}=1$ for all $\phi$.

\subsection{Phase from the Rotated Basis}
For the orthonormal analyzer Eq.~\eqref{eq:orthobasis}, the two-port per-event FI can be written compactly by defining
\begin{align*}
    A &= T\cos^{2}\theta+\sin^{2}\theta, \\
    B &= T\sin^{2}\theta+\cos^{2}\theta, \\
    \Delta &= 2\sqrt{T}\sin\theta\cos\theta\cos\phi,
\end{align*}
as
\begin{equation}
    \mathcal{I}_{\phi}^{(\theta)}=\frac{4T\sin^{2}\theta\cos^{2}\theta\,\sin^{2}\phi}{T+1}
    \left[\frac{A}{A^{2}-\Delta^{2}}+\frac{B}{B^{2}-\Delta^{2}}\right]\!,
    \label{eq:IF_phi_theta_general}
\end{equation}
which reduces at $T=1$ to
\begin{equation}
\mathcal{I}_{\phi}^{(\theta)}\big|_{T=1}=\frac{\sin^{2}(2\theta)\,\sin^{2}\phi}{1-\sin^{2}(2\theta)\,\cos^{2}\phi}\,,
\label{eq:IF_phi_theta_T1}
\end{equation}
maximized at $\theta=\pi/4$.

\subsection{Single-Port vs.\ Two-Port}
For D/A at any $T$, the {per-used-event} FI for a single port equals Eq.~\eqref{eq:IF_phi_DA}. Two-port processing therefore {doubles the information rate per detected pair} (it uses all coincidences) and yields a balanced normalized estimator [Eqs.~\eqref{eq:V_DA}, \eqref{eq:V_theta}]. Furthermore, the visibility of image acquisition on rotational $|e_{\pm}(\theta)\rangle$ bases at $\theta=\pi/8$ is shown in Fig. \ref{fig:visibility} (c).

\subsection{Phase Retrieval from $\cos\phi$}
Eq.~\eqref{eq:V_DA} and~\eqref{eq:V_theta} provide $\cos\phi(x,y)$. To recover $\phi$ unambiguously one needs: (i) phase stepping (apply a known global dither $\delta$ on one interferometer arm during the same run and jointly fit $\cos\phi$ and $\cos(\phi+\delta)$), (ii) a brief second calibration run with a known phase offset, or (iii) spatial unwrapping with smoothness priors where valid. To illustrate phase retrieval, the estimated phase (mod $\pi$) using Eqs.~\eqref{eq:V_theta} at $\theta=\pi/8$ is shown in Fig. \ref{fig:visibility} (d).

\section{Information-Theoretic Analysis}
For small parameter changes $\delta\boldsymbol{\vartheta}$ with $\boldsymbol{\vartheta}=(T,\phi)^{\!\top}$, local distinguishability is set by $D_{\mathrm{KL}}\!\approx\!\tfrac{1}{2}\,\delta\boldsymbol{\vartheta}^{\!\top}\mathcal{J}\,\delta\boldsymbol{\vartheta}$, where the total FIM is
\begin{equation}
\mathcal{J} = S_T\,\mathcal{I}^{(\mathrm{WPI})} + S_\phi\,\mathcal{I}^{(\mathrm{erase/rot})},
\end{equation}
with $\mathcal{I}^{(\mathrm{WPI})}=\mathrm{diag}\!\big(\tfrac{1}{T(1+T)^{2}},\,0\big)$ and $\mathcal{I}^{(\mathrm{erase/rot})}$ from Eqs.~\eqref{eq:IF_phi_DA} or \eqref{eq:IF_phi_theta_general}.

\subsection{Equivalence to Time-Division}
\textbf{Proposition 1 (FIM equivalence under labeled randomization).} Consider QEI where each coincidence is recorded along with the ancilla basis/outcome, and the random choice over bases $\{\theta_k\}$ is independent of $(T,\phi)$. Assume equal detection efficiencies across B outcomes and identical object-arm configuration, negligible drift and accidentals (or appropriately corrected). Then the {per-coincidence} FIM of QEI equals the convex combination of the per-coincidence FIMs of time-division experiments that run each basis $\theta_k$ for fraction $w_k$:
\[
\mathcal{J}_{\mathrm{QEI}}=\sum_k w_k\,\mathcal{J}(\theta_k).
\]
The observed outcomes include the label $(\theta_k,\text{outcome})$, so the log-likelihood decomposes as $\sum_i \log p(y_i|\theta_{k_i};\boldsymbol{\vartheta})$. Taking expectations under the product measure with fixed $\{w_k\}$ gives $\mathbb{E}[-\partial_{\boldsymbol{\vartheta}}^{2}\log\mathcal{L}]=\sum_k w_k\,\mathbb{E}[-\partial_{\boldsymbol{\vartheta}}^{2}\log p(\cdot|\theta_k)]$, yielding the stated equality. 

\subsection{Operational Advantage}
By operational, we mean advantages in experimental workflow and implementation, rather than enhancements to per-photon information: (i) single-run acquisition of both modalities without touching the object arm; (ii) perfect co-registration since both reconstructions derive from the same time-tagged acquisition stream, partitioned into basis-labeled subsets; (iii) remote/delayed basis choice (and continuous tuning) via the ancilla measurement, reducing drift/alignment risk relative to time-division.

\section{Discussion and Conclusion}
QEI is useful in two ways: it helps make the physics intuitive, and it gives a workable imaging protocol. Conceptually, the orthonormal analyzer sweep [Eqs.~\eqref{eq:orthobasis}--\eqref{eq:V_theta}] turns the abstract complementarity trade-off into a pixel-resolved morphing of images while respecting completeness/no-signaling [Eq.~\eqref{eq:singles_nosignal}]. Technically, the explicit estimators and FI/CRBs quantify performance and clarify that QEI's advantages are operational: one run, co-registration, and a remote, even delayed, modality choice.

We presented a theory of QEI with: (i) an {orthonormal} intermediate analyzer enabling a continuous trade-off, (ii) closed-form dual-modality estimators and visibilities [Eqs.~\eqref{eq:T_estimator_twoports}, \eqref{eq:V_DA}, \eqref{eq:V_theta}], (iii) singles/no-signaling checks, and (iv) FI/CRBs [Eqs.~\eqref{eq:IF_phi_DA}--\eqref{eq:IF_phi_theta_T1}] and a precise statement of information-theoretic equivalence to time-division. These results sharpen QEI as both a clean visualization of complementarity and a convenient imaging protocol.

\onecolumngrid
\appendix

\section{QEI and Related Imaging}
\begin{table}[htbp]
\centering
\scriptsize
\setlength{\tabcolsep}{3pt}
\renewcommand\theadfont{\bfseries}

\begin{tabularx}{\textwidth}{@{}>{\raggedright\arraybackslash}p{2.9cm}
                          >{\raggedright\arraybackslash}p{2.9cm}
                          >{\raggedright\arraybackslash}p{2.9cm}
                          >{\raggedright\arraybackslash}p{2.9cm}
                          >{\raggedright\arraybackslash}p{2.9cm}
                          >{\raggedright\arraybackslash}p{2.9cm}@{}}
\toprule
\thead{Capability}
& \thead{QEI \\ (this work)}
& \thead{Imaging with \\ Undetected Light \\ \cite{Lemos2014} }
& \thead{Phase-Quadrature \\ QIUP \\ \cite{Haase2022}}
& \thead{Entangled \\ Holography \\ \cite{Defienne2021}}
& \thead{QCPGM \\ \cite{PosMom2024}} \\
\midrule
Quantum resource & Entangled pairs (pol.) & Induced coherence & Induced coherence + pol. & Spatial/pol. ent. pairs & Pos-mom entangled pairs \\
Primary readout & Ancilla-sorted coinc. & Direct intensity & 2-channel direct intensity & Correlations (hologram) & Near/far-field coinc. \\
Modality outputs & T \& $\phi$ (same data) & T or phase (reconfig.) & Phase \& Vis. (\pmark for T) & Complex field (T \& $\phi$) & T \& phase (via momentum) \\
Single-acquisition & \cmark & \xmark & \cmark (two channels) & \cmark (one dataset) & \cmark \\
Remote/delayed choice & \cmark (ancilla choice) & \pmark (toggle path) & \xmark (fixed encoding) & \xmark (fixed readout) & \xmark (fixed readout) \\
Continuous trade-off & \cmark (rotatable $\theta$) & \xmark (binary) & \xmark (fixed mixing) & \xmark (fixed) & \xmark (fixed) \\
Two-port normalization & \cmark ($\theta$-independent) & \xmark & \xmark & \xmark & \xmark \\
FI/CRB analysis & \cmark (T, $\phi$) & \xmark & \pmark (metrics) & \pmark (fidelity) & \pmark (precision) \\
Object arm unchanged & \cmark & \xmark (toggle) & \cmark & \cmark & \cmark \\
Perfect co-registration & \cmark (post-sorted) & \xmark (separate runs) & \cmark (simultaneous) & \cmark (same data) & \cmark (paired events) \\
2D scenes/samples & \cmark & \cmark & \cmark & \cmark & \cmark \\
ICO / causal-witness & \cmark (optional) & \xmark & \xmark & \xmark & \xmark \\
Operational advantage & Single-run, co-reg. & Undetected photons & Single-shot phase & Remote/incoherent phase & Background-resilient QPI \\
\bottomrule
\end{tabularx}

\caption{\textbf{Comparison of QEI with closely related quantum/entanglement-based imaging approaches.}
Legend: \cmark\ = supported; \xmark\ = not supported; \pmark\ = partial/conditional.
``Dual modality'' means recovering both absorption $T(x,y)$ and phase $\phi(x,y)$ (or equivalent contrast/phase) from a single acquisition.
QEI uniquely combines: (i) single-run, co-registered $T$ and $\phi$ via retrospective ancilla sorting; (ii) a {continuous} analyzer trade-off
with two-port, analyzer-independent normalization; (iii) explicit FI/CRB analysis. 
}
\label{tab:method_comparison}
\end{table}


\nocite{*}
\twocolumngrid
\bibliography{references}

\end{document}